# An approximate cooperativity analysis by DSC of phase transitions of DPPC-DOCNa dispersions

ŞTEFAN HOBAI
*Department of Biochemistry, University of Medicine and Pharmacy, Tg. Mureş, Romania*

**Abstract**
The effect of solute, sodium deoxycholate bile salt (DOCNa), on the cooperativity of ordered-fluid fosfolipid phase transition has been investigated by determining of the transition widths in multibilayer dispersions of the dipalmitoylphosphatidylcholine (DPPC)-DOCNa mixtures. The $\Delta T_{1/2}$ values were determined by differential scanning calorimetry. The van't Hoff values of the main transition enthalpies were calculated using an approximative expression, deduced from Zimm and Bragg theory [Sturtevant,J.M.(1982), Proc.Natl.Acad.Sci.USA,79:3963-3967]. The bile salt decreases the cooperativity (size of cooperative unit) of DPPC bilayers.

**Introduction**

Phenomenological theories of the cooperativity of phase transitions have been formulated on the basis of coexisting cluster of lipid molecules, characteristic of the different phases, at the phase transition. The Zimm and Bragg theory of cooperative phase transitions supposes that during the transition there are different states of lipid. Considering three states, ordered, fluid and interfacial states [Marsh,1976 cit.in: Marsh,1977], the degree of transition, $\theta$, i.e. mean fraction of molecules in the fluid state, is:

$$\theta = \frac{1}{2}\left[1 + \frac{s-1}{\sqrt{(s-1)^2 + 4\sigma s}}\right] \quad (1)$$

where $s$ is the statistical weight for the state above the transition and $\sigma$ is the cooperativity parameter. The values of $s$ and $\sigma$ are related with the free energies of fluid and interfacial states by the arrhenian exponentials:

$$s = \exp(\Delta S_t/R)\exp(-\Delta H_t/RT) \quad (2)$$

$$\sigma = \exp(S_i/R)\exp(-H_i/RT) \quad (3)$$

where $\Delta H_t$ and $\Delta S_t$ are calorimetric transition enthalpy and entropy, respectively. $H_i$ and $S_i$ are interfacial enthalpy and entropy, respectively.

Cooperativity arises from interfacial free energy, $G_i = H_i - TS_i$. The chain-melting transitions of saturated phosphatidylcholines are highly cooperative [Marsh,1991] so that it becomes indistinguishable from a first-order, discontinuous transition [Lee,1977]. Differentiating eqn. (1):

$$\left(\frac{d\theta}{dT}\right)_{T_t} = \frac{1}{4\sqrt{\sigma}} \cdot \frac{\Delta H_t}{RT_t^2} \tag{4}$$

shows that $\theta$ will have a linear dependence on $1/T$ around the transition temperature, $T_t$, with a negative slope from which $\sigma$ may be determined. The values of $\theta$ were calculated from the ESR and rectangular light-scattering measurements [Marsh,1977].

If the ordered-fluid transition is considered as a pseudo-unimolecular reaction with equilibrium constant $K_t = \theta/(1-\theta)$, eqn. (2) can be transformed into the van't Hoff (vH) form:

$$\left(\frac{d\theta}{dT}\right)_{T_t} = \frac{\Delta H_{vH}}{4RT_t^2} \tag{5}$$

Equalization member to member of eqns. (4) and (5) produces:

$$\sigma = (\Delta H_t / \Delta H_{vH})^2 \tag{6}$$

The smaller the value of $\sigma$, the greater the cooperativity of transition. The quantity $n=1/\sqrt{\sigma}$ is referred to as the size of the *cooperative unit* and is directly related to the domain size, being equal to the mean number of lipids per interfacial lipid within a domain at the centre of the transition. The pH-dependence of the phase transition of DMPA has been investigated using DSC [Blume,1979]. The high cooperativity at pH 3.5 indicated that a tightly packed structure is formed, which increases the size of the cooperative unit.

A matter of large interest was the influence of small molecules, as anesthetics, on the thermodynamic parameters of thermotropic behaviour of saturated phosphatidylcholines. Procaine reduces the cooperativity of the main phase transition of DMPC bilayers [Tsong,1977] and a cluster model suggests that the anesthetic reduces the average size of the lipid clusters.

If the lipid molecules in the bilayers would be perfectly ordered the transition would be isotermal. However, there are many limitations generated by experimental conditions which lead to the extension of observed transitions over finite ranges of temperature. The reasons of this broadening can be the imperfect ordering of the lipid molecules in the bilayer, finite scan rates and calorimetric lags, and the presence of traces of impurities.

The experimental DSC curve is the temperature rate of heat absorption, equivalent to the excess heat capacity function, $C_{ex}$, given by eqn.:

$$C_{ex} = \Delta H_t \cdot d\theta/dT \tag{7}$$

The sharpness of the $C_{ex}$ peak of main phase transition emphasizes the cooperativity of it. The apex of the curve corresponds to transition temperature, $T_t$, and to maximal value of

$$C_{ex,max} = \Delta H_t \cdot \left(\frac{d\theta}{dT}\right)_{T_t} \quad (8)$$

With certain assumptions was developed [Mastrangelo,1955 cit.in Sturtevant,1982] an expression for the broadening of the main transition of a pure lipid caused by a solute forming ideal solutions in both gel and fluid phases. During the phase transition, the equilibrium in distribution of solute between the lipid phases is maintained. This equilibrium and the partitioning of solute between aqueous and lipid phases do not change appreciably within the temperature range of the transition. The bilayer behaves as a isotropic solvent. In these conditions the eqn. (4) in eqn. (8) may be substituted, resulting a relation between $\Delta H_{vH}$ and $\Delta H_t$ for a "pure" lipid:

$$C_{ex,max} = \Delta H_t \Delta H_{vH} / 4RT_0^2 \quad (9)$$

Here $T_0$ is the temperature at which the phase transition of the "pure" lipid (lipid containing no added solute) is half completed.

$\Delta H_{vH}$ may be expressed by less exact formulae, derived from eqn. (9):

$$\Delta H_{vH} \approx 6.9\, T_0^2 / \Delta T_{1/2} \quad (10)$$

where $\Delta T_{1/2}$ is the width (in degrees) at half-height of the transition. $\Delta T_{1/2}$ is related to both the cooperative nature of the transition and the purity of the system. Presence in the bilayers of impurities or solute molecules reduces lipid-lipid interactions and, consequently, reduces the cooperativity of phase transitions.

**Chemicals**

Dipalmitoylphosphatidylcholine(DPPC) was purchased from Sigma Chemical Co. DPPC was shown to be pure by thin-layer chromatography on Silicagel using a solvent system of chloroform: methanol:water (65:25:4; v/v) with iodine staining. DOCNa was purchased from Merck and TRIS was purchased from Austranal. Both compounds were used without further purification. Organic solvents were of analytical grade. Double-distilled water (glass still) was used to make all samples.

**Solutions, dispersions and calorimetric measurements**

The synthetic lipid DPPC and the bile salt DOCNa were solved at different molar ratios in 0.3 ml ethanol at 20°C (Table 1). The evaporation of the solvent was made under nitrogen stream. MLV solutions were obtained by hydration with 40µl Tris-Cl 0.05M, pH=7.2. The samples were subjected to DSC analysis, with calorimeter type Du Pont Instruments at 5°C/min heating speed. The temperatures of phase transitions and the transition width, $\Delta T_{1/2}$ are shown in Table 1.

Starting from ethanolic solutions of DPPC and DOCNa respectively, which were mixed at different molar ratio, there were prepared dry films having the compositions specified in Table 2. Hydration of these films was made by vortexing with a buffer Tris-Cl 0.01M, pH=7.4 at a 0.66mg lipid concentration, after which the MLV-DPPC-DOCNa suspensions were left at rest overnight. DSC measurements with calorimeter type VP-DSC model MN2State produced thermograms whose thermodynamic caracteristics are mentioned in Table 2.

**Results and discussion**

The graphical determination of ΔH value was made by measuring the area under the peak of excess heat capacity. For example:

$\Delta H_p^{measured}$ = 1,36mJ/mg = 998,2 J/mol;

$$\Delta S_p^{calculated} = \frac{\Delta H_p^{measured}}{T_p} = \frac{998,2}{308,91} = 3,23 \text{ J/mol.K} \qquad (11)$$

where $T_p = t_p$ (°C) +273,16 = 35,75 +273,16 =308,91 K.

$\Delta H_t^{measured}$ = 19,77mJ/mg = 14511 J/mol;

$$\Delta S_t^{calculated} = \frac{\Delta H_t^{measured}}{T_t} = \frac{14511}{314,41} = 46,15 \text{ J/mol.K} \qquad (12)$$

where $T_t = t_t$ (°C) +273,16 =41,25+ 273,16 = 314,41 K.

As it can be observed in Table 1 DOCNa has strong effect on the caloric peak of the main phase transition. DOCNa significantly decreases $T_m$ value, proportionally to its molar fraction. The decrease of $T_m$ value is about 5 degrees in case DPPC:DOCNa=2:1 suggests that solid solutions play an important role in this system. A major increase of $\Delta T_{1/2}$ can be observed as a result of the decrease of phase transition cooperativity.That is determined by the penetration of DOCNa molecules in phospholipid bilayers, determining the decrease of the van der Waals interactions.

**Table 1.** *Temperatures of phase transitions and transition widths of DPPC and mixtures DPPC:DOCNa at molar ratios 1:2 and 2:1 obtained by Du Pont Instruments equipment.*

| Composition DPPC:DOCNa in dry film | $T_p$ (°C) | $T_t$ (°C) | $\Delta T_{1/2}$ (°C) |
|---|---|---|---|
| DPPC | 36.0±0.25 | 41.5±0.3 | 1.42±0.14 |
| 10:1 | - | 40.83±0.38 | 2.03±0.4 |
| 2:1 | - | 36.07±0.12 | 3.07±0.12 |

DOCNa abolishes the caloric effect of the pretransition. At a DPPC:DOCNa 1:2 ratio, the caloric effect of the phase transition can be no longer emphasized, probably due to the solubilization of phospholipid vesicles.

**Table 2.** *Thermotropic and thermodynamic characteristics of MLV-DOCNa systems equilibrated with buffer Tris-Cl 0.01M, pH=7.4, obtained by VP-DSC MN2 State equipment.*

| | Composition DPPC:DOCNa in dry film | $T_p$ (°C) | $\Delta H_p$ (kJ/mol) | $T_t$ (°C) | $\Delta H_m$ (kJ/mol) | $\Delta T_{1/2}$ (°C) |
|---|---|---|---|---|---|---|
| 1 | DPPC | 35.75 | 1.806 | 41.25 | 26.259 | 0.75 |
| 2 | 10:1 | 31.72 | 1.818 | 40.91 | 27.546 | 0.5 |
| 3 | 5:1 | 24.03 | 0.539 | 40.58 | 36.742 | 0.78 |
| 4 | 1:1 | - | - | 41.23 | 34.665 | 3.2 |

The values of $T_p$ and $T_m$ decrease and the values of $\Delta H_p$ and $\Delta H_m$ increase up to the dry mixture molar ratio no.3. The system produced by hydration of the mixture no.4 is probably in micellar phase. The $\Delta T_{1/2}$ value also indicate a destabilization of the lamellar phase in the case of suspension no.4. This micellisation phenomenon takes place in the course of absorption in the digestive tube a process which is favoured by bile salts. Mixed micelles mediate the intestinal absorption of some products of lipid digestion.

The values of the cooperativity parameter, σ and the sise of the cooperative unit of the main phase transition of DPPC-DOCNa systems, CU, are included in Table 3. The van't Hoff enthalpy, $\Delta H_{vH}$, was calculated with an equation (10) and the calorimetric enthalpy, $\Delta H_t$, was measured both with DSC Du Pont Instruments and with VP-DSC type MN2State.

**Table 3.** *The values of the cooperativity parameter, σ, and the size of the cooperative unit, CU, of main phase transition of DPPC-DOCNa systems.*

| Composition (device) | $\Delta H_{vH}$ (kJ/mol) | $\Delta H_t$ (kJ/mol) | $\sigma \cdot 10^3$ | CU |
|---|---|---|---|---|
| DPPC (*Du Pont Instruments*) | 909.453 | 14.511 | 0.255 | 62.7 |
| *DPPC(VP-DSC)* | 909.453 | 26.259 | 0.833 | 34.6 |
| DPPC:DOCNa (10:1,mol) | 1361.231 | 27.546 | 0.41 | 49.4 |
| DPPC:DOCNa (5:1,mol) | 870.752 | 36.742 | 1.78 | 23.7 |
| DPPC:DOCNa (1:1,mol) | 213.126 | 34.665 | 26.45 | 6.1 |

The cooperativity parameter is a square ratio of measured calorimetric enthalpy and calculated enthalpy (eqn.6).:

The size of the cooperative unit (CU) is calculated on the basis of the equation:

$$CU = 1/\sqrt{\sigma} \qquad (13)$$

The values presented in Table 3 show the decrease of main phase transition cooperativity of the DPPC-DOCNa bilayers directly proportional to the increase of the surfactant concentration amount.

**Abbreviations**

$c_{ex}$ – excess heat capacity
$c_{ex,max}$ – maximum value of excess heat capacity

CU - size of cooperative unit

DOCNa-sodium deoxycholate

DPPC-dipalmitoylphosphatidylcholine(di-$C_{16}$PC)

DSC-differential scanning calorimetry

$G_i$ – interfacial free energy

$\Delta H_t$ - calorimetric transition enthalpy

$H_i$ – interfacial enthalpy

$\Delta H_p$ - calorimetric pretransition enthalpy

$\Delta H_{vH}$ - van't Hoff transition enthalpy

*s* - statistical weight for the state above the transition

$\Delta S_t$ – calorimetric transition entropy

$S_i$ – interfacial entropy

$\Delta S_p$ - pretransition entropy

$T_p$ - pretransition temperature

$T_t$ - main phase transition temperature

t – temperature
T- absolute temperature

$T_0$ – phase transition temperature of the "pure" phospholipid

$\Delta T_{1/2}$ - transition width

σ - cooperativity parameter

*θ* - mean fraction of molecules in the fluid state

**Correspondence address:**

Ştefan Hobai, Departament of Biochemistry, University of Medicine and Pharmacy, Târgu Mureş, Romania. Fax: 040-065-210407, E-mail: fazy@netsoft.ro